\documentclass[useAMS,usenatbib]{mn2e} 
\usepackage{graphics,graphicx} 
\title[An investigation of the absolute circular polarization in radio
pulsars] {An investigation of the absolute circular polarization in radio
pulsars} 
\author[Karastergiou \& Johnston]
{A.~Karastergiou$^{1,2}$ and S.~Johnston$^1$ \\
$^1$School of Physics, University of Sydney, NSW 2006, Australia\\
$^2$Max-Planck Institut f\"ur Radioastronomie, Auf dem H\"ugel 69,
53121 Bonn, Germany} 
\date{Released 2004 Xxxxx XX}
 
\pagerange{\pageref{firstpage}--\pageref{lastpage}} \pubyear{2004} 
 
\def\LaTeX{L\kern-.36em\raise.3ex\hbox{a}\kern-.15em 
    T\kern-.1667em\lower.7ex\hbox{E}\kern-.125emX}

\begin{document} 
 
\label{firstpage} 
 
\maketitle 
 
\begin{abstract} 
In most pulsars, the circularly polarized component, Stokes $V$, is
weak in the average pulse profiles. By forming the average profile of
$|V|$ from single pulses we can distinguish between pulsars where $V$
is weak in the individual pulses and those where large $V$ of variable
handedness is observed from one pulse to the other. We show how $|V|$
profiles depend on the signal-to-noise ratio of $V$ in the single
pulses and demonstrate that it is possible to simulate the observed,
broad distributions of $V$ by assuming a model where $|V|$ is
distributed around a mean value and the handedness of $V$ is permitted
to change randomly. The $|V|$ enhanced profiles of 13 pulsars are
shown, 5 observed at 1.41 GHz and 8 observed at 4.85 GHz, to complement
the set in Karastergiou et al. (2003b). It is argued that the degree
of circular polarization in the single pulses is related to the
orthogonal polarization mode phenomenon and not to the classification
of the pulse components as cone or core.
\end{abstract} 
 
\begin{keywords} 
pulsars: general - polarization
\end{keywords} 
\section{Introduction}
Despite much progress in understanding the polarimetric properties of
pulsar radio emission, there remains a number of open and, in some
cases, unexplored questions. The basic knowledge that has been
obtained over the years is mainly related to three observed
phenomena. The first is the observation made by Radhakrishnan \& Cooke
(1969\nocite{rc69a}), that the average polarization position angle
(PA) swings across the pulse profile in a way that can be explained by
the geometrical {\it rotating vector model}, in which the position
angle is tied to the direction of the magnetic field lines. As the
emission beam of the pulsar sweeps past the line of sight of the
observer, an observed swinging PA is a natural consequence. This model
has been used to derive the key geometrical angles for a large number
of pulsars, such as the inclination angle of the magnetic to the
rotational axis and the impact parameter, which determines how close
the magnetic axis passes to the line of sight.

The second phenomenon is related to orthogonal transitions observed in
the PA of single pulses and integrated pulse profiles. The observation
of Manchester, Taylor \& Huguenin (1975\nocite{mth75}) that the PA at
a particular pulse phase was often observed to be orthogonal to its
common value, was taken one step further by Backer, Rankin \& Campbell
(1976\nocite{brc76}) and Cordes, Rankin \& Backer
(1978\nocite{crb78}), who identified these PA variations as the
manifestation of orthogonal polarization modes (OPM) of emission,
anti-parallel to each other on the Poincar\'{e} sphere. The link
between the PA jumps and the handedness of circular polarization
($V$), noticed by Cordes et al. (1978), was essential to this
achievement. The early single-pulse observations permitted the study
of the PA at particular pulse longitudes and showed that OPM jumps in
single pulses are accompanied by reduced linear polarization $L$, an
indication that the observed radiation is in fact the incoherent sum
of the emitted OPMs. Stinebring et al. (1984\nocite{scr+84}) developed
a model in which the polarization properties can be attributed to
statistical fluctuations in the OPMs. This model was developed further
by McKinnon \& Stinebring (1998\nocite{ms98}, 2000\nocite{ms00}) and
has been recently extended to account for deviations from mode
orthogonality (McKinnon 2003a\nocite{mck03b}). Meanwhile, Karastergiou
et al. (2001, 2002\nocite{khk+01}\nocite{kkjl+02}) showed that the OPM
phenomenon was broadband in single pulses, simultaneously observed at
different frequencies. Recently, Karastergiou et
al. (2003\nocite{kjk03}) showed that the association between the
handedness of $V$ and the PA becomes weaker towards higher
frequencies, indicating that the observed circular polarization cannot
straightforwardly be accounted for by existing OPM models.

The third and mostly unexplained observational fact is the high degree
of circular polarization in pulsar emission itself.  In single pulses,
the fractional circular polarization is often quite high, despite the
fact that, on average, it usually does not exceed $10\%$ (Han et
al. 1998)\nocite{hmxq98}. Radhakrishnan \& Rankin (1990) noticed that
the average circular polarization observed in central or {\it core}
components was generally greater than that in the outer or {\it cone}
components. Another feature of core components is that the integrated
$V$ profile is often seen to swing between one and the other
handedness. There was a claim that the direction of this change was
related to the direction of the swing of the PA (also by Radhakrishnan
\& Rankin 1990\nocite{rr90}), however, this claim was contested by Han
et al. (1998)\nocite{hmxq98} and Gould (1990).

In this paper, we document the statistics of $|V|$ and provide
guidelines for the interpretation of $|V|$ profiles. This is done by
examining the dependence of the $|V|$ profile on the distribution of
$V$ in the single pulses. Then we proceed to show a number of $|V|$
enhanced integrated profiles from recent polarization observations and
finally we discuss the associations of $V$ with OPMs and especially
whether there are differences between core and cone emission.

\section{Single-Pulse Statistics}

The phase-resolved intensity (Stokes $I$) distribution of a number of
pulsars has been studied by Cairns et al (2001,
2004)\nocite{cjd01,cjd04}.  They showed that the statistics are
characterised by a log-normal distribution, i.e.  that a histogram of
the logarithm of the flux densities is normally distributed.
Typically, the standard deviation of the distribution is less than 0.3
in the log (see e.g. Johnston et al. 2001, Cairns et al. 2004,
Johnston 2004) \nocite{jvkb01,cjd04,joh04} consistent with the
observational evidence that very few pulsars have single pulses with
energy more than 10 times the mean energy.

In a study of the Vela pulsar, Kramer, Johnston \& van Straten (2002)
\nocite{kjv02} showed that the distribution of $V$ was strongly
correlated with the total intensity distribution. The correlation was
not perfect; the distribution of $V/I$ is a Gaussian with the standard
deviation determined from both instrumental noise and seemingly
intrinsic random fluctuations in $V$. These fluctuations were
sometimes large enough that the handedness of the circular
polarization changed even though the same orthogonal mode dominated
the total intensity.  However, the result for Vela is consistent with
the earlier, low frequency measurements by Cordes et al. (1978). They
showed that the sign of the circular polarization is highly correlated
with the mode of the dominant emission; a change of the dominant mode
caused a change in sign of circular polarization.

These observations fit in well with the preferred model of orthogonal
modes as outlined by McKinnon \& Stinebring (1998)\nocite{ms98}. In
the model, the orthogonal modes occur simultaneously and are 100\%
polarized. Each mode is associated with a particular handedness of
circular polarization.  The distribution of $I$ is the
result of the sum of the two orthogonal modes and the distribution of
$V$ results from the difference of the two modes, as does the
distribution of $L$. 
If the fluctuations of the two modes are perfectly covariant in time,
then the difference between them will be constant. This means that the
observed degree of $L$ and $V$ will remain constant and no PA jumps
will occur. On the other hand, if the covariance between the mode
intensities were low, this would result in a variety of observed
polarization states. The difference in the OPM intensities would then
give rise to broad distributions in $L$ and $V$. Furthermore,
non-covariant fluctuations in the OPM intensities can result in
switching of the dominant OPM, which is observed as a $90^o$ jump in
the PA. Therefore, the polarization properties observed in pulsars
according to the superposed OPM model, strongly depend on both the
mean intensities of the two OPMs and on their covariance.

Recently, however, Karastergiou et al. (2003a)\nocite{kjk03},
performed high frequency observations of PSR~B1133+16. Surprisingly,
they discovered that very large values of circular polarization with
either sign were present in the data, and that {\bf the sign of the
circular polarization was not well correlated with the dominant
orthogonal mode}.  Furthermore, in observations of PSR~B0329+54,
Karastergiou et al. (2001)\nocite{khk+01} showed that the outer
components of the profile had significant values of circular
polarization in single pulses but the distribution was such that the
resultant integrated profile had virtually no circular polarization
present.

We note that all these studies have been done for bright pulsars with
high signal to noise ratios in individual pulses. However, in the
majority of pulsars the signal to noise is generally low, especially
for circular polarization. This issue was tackled by Karastergiou et
al. (2003b, hereafter KJMLE)\nocite{kjm+03} who used single pulses to
construct a profile of the absolute value of circular polarization,
$|V|$. They found that the single-pulse behaviour of $V$ is affected,
as expected, by OPM jumps, that there are particular pulse phases with
significant total power emission that have constant, near zero,
circular polarization and that the changes in frequency of $|V|$
profiles are much less significant than the changes in $V$ profiles,
indicating that $|V|$ is an important quantity in pulsar emission. 

\begin{figure*}
\centerline{
\resizebox{\hsize}{!}{\includegraphics[angle=-90]{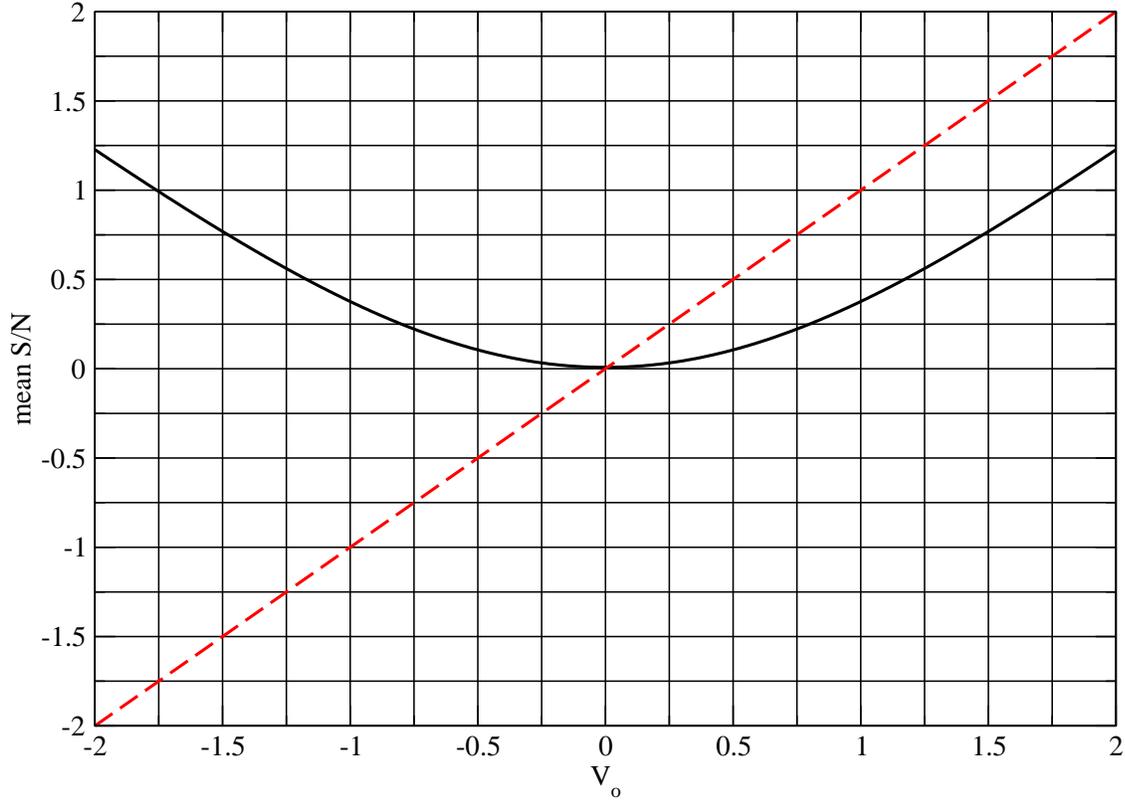}} }
\caption{The average S/N in $V$ (dashed) and $|V|$ (solid) is shown
  against a constant $V_o$ in the single pulses, as given by Eq. 1,
  with $\sigma_n=1$. The final S/N obtained by integrating $N$ pulses
  is $\sqrt{N}$ times the y-axis value.}
\label{snr}
\end{figure*}

\section{The statistics of $|V|$}

\begin{figure*}
\centerline{
\resizebox{0.5\hsize}{!}{\includegraphics[angle=-90]{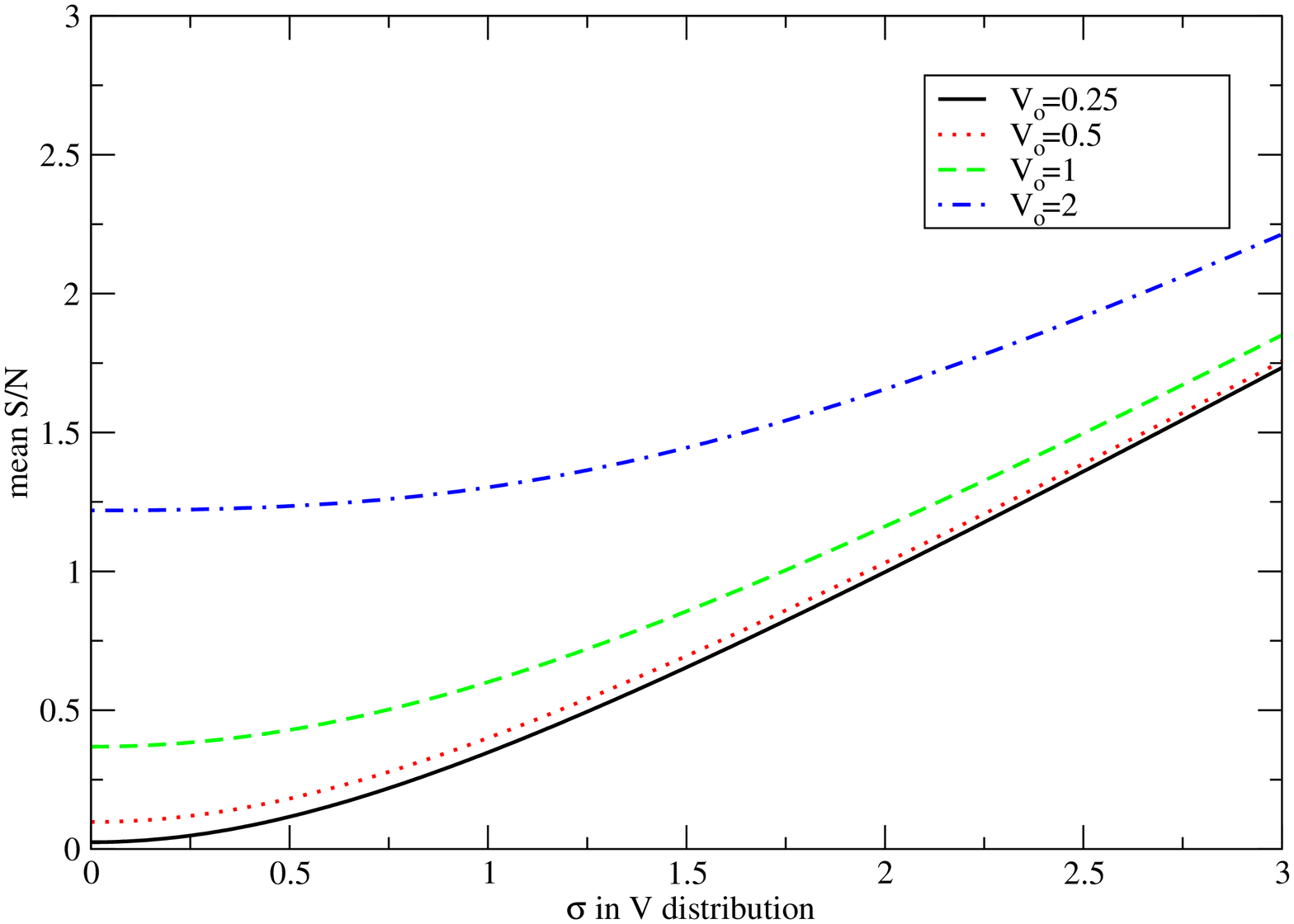}} 
\resizebox{0.5\hsize}{!}{\includegraphics[angle=-90]{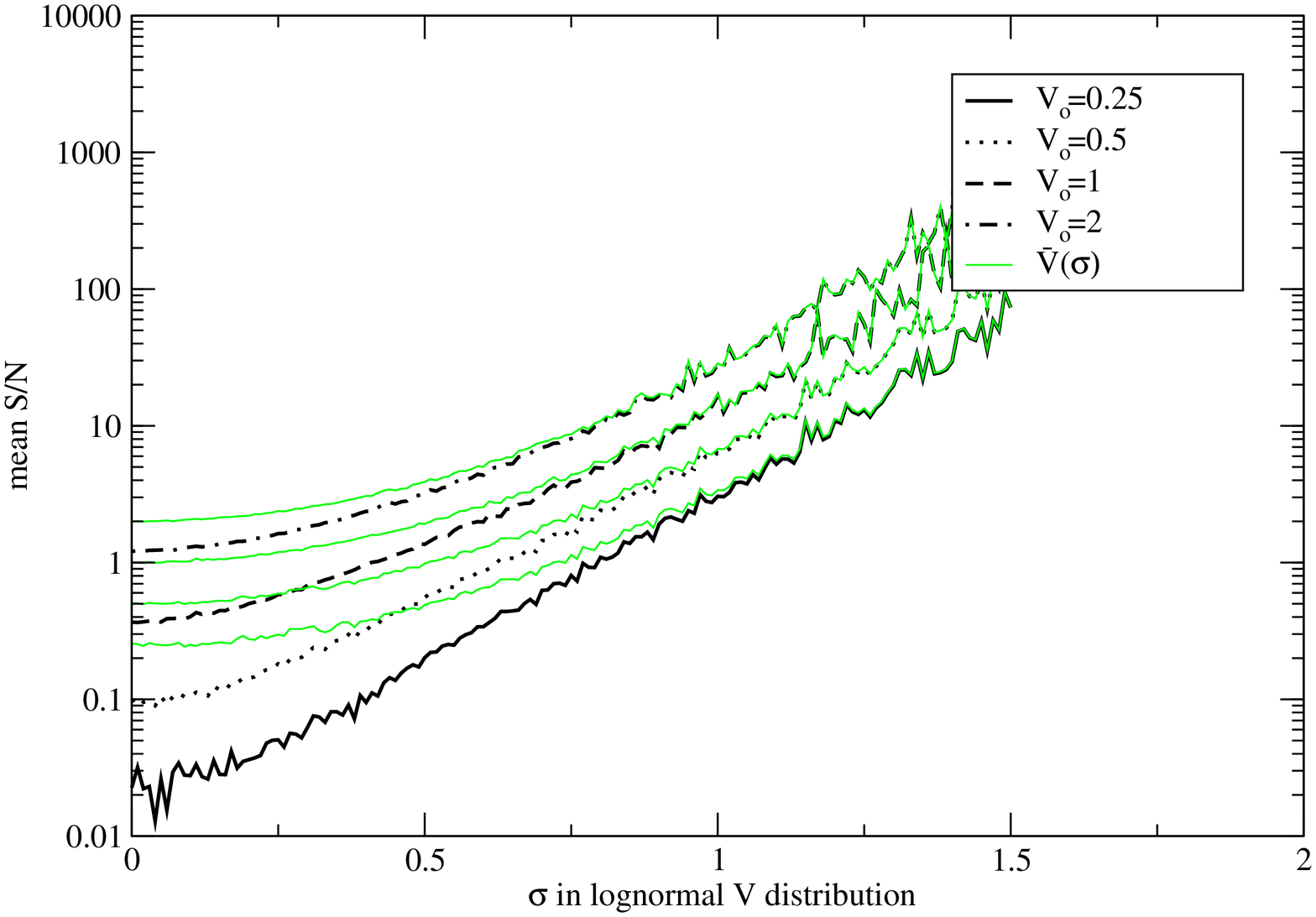}} 
}
\caption{Left panel: The mean S/N in $|V|$, computed from the analytic
expression in equation 2, is shown against the $\sigma$ in a Gaussian
distribution of $V$, for four values of $V_o$.  Right panel: The mean
S/N in $|V|$ computed from simulations of a log-normal distribution in
$V$ for four values of $V_o$.  Also shown is the mean S/N in $V$. In
both panels $\sigma_n=1$}
\label{sigma}
\end{figure*}

\begin{figure*}
\centerline{
\resizebox{\hsize}{!}{\includegraphics[angle=-90]{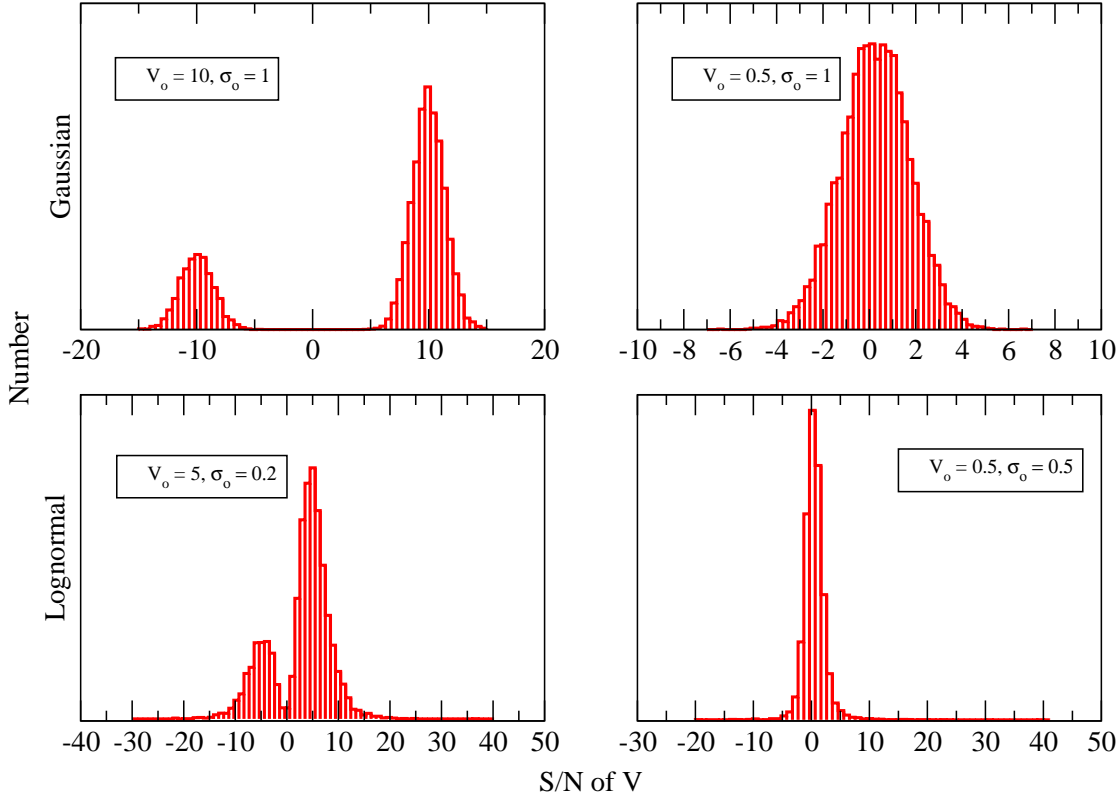}}
}
\caption{Four simulated distributions of $V$ in single pulses. Each simulation
is constructed from an intrinsic $V$ distribution which is bi-modal
convolved with Gaussian (receiver) noise. The intrinsic distributions
consist of two modes of opposite mean and equal $\sigma$ as denoted;
the top panels arise from a Gaussian distribution and the bottom
panels from a log-normal distribution. In all cases there are
three times as many pulses with positive sign as negative.}
\label{distributions}
\end{figure*}

\begin{figure*}
\centerline{
\resizebox{\hsize}{!}{\includegraphics[angle=-90]{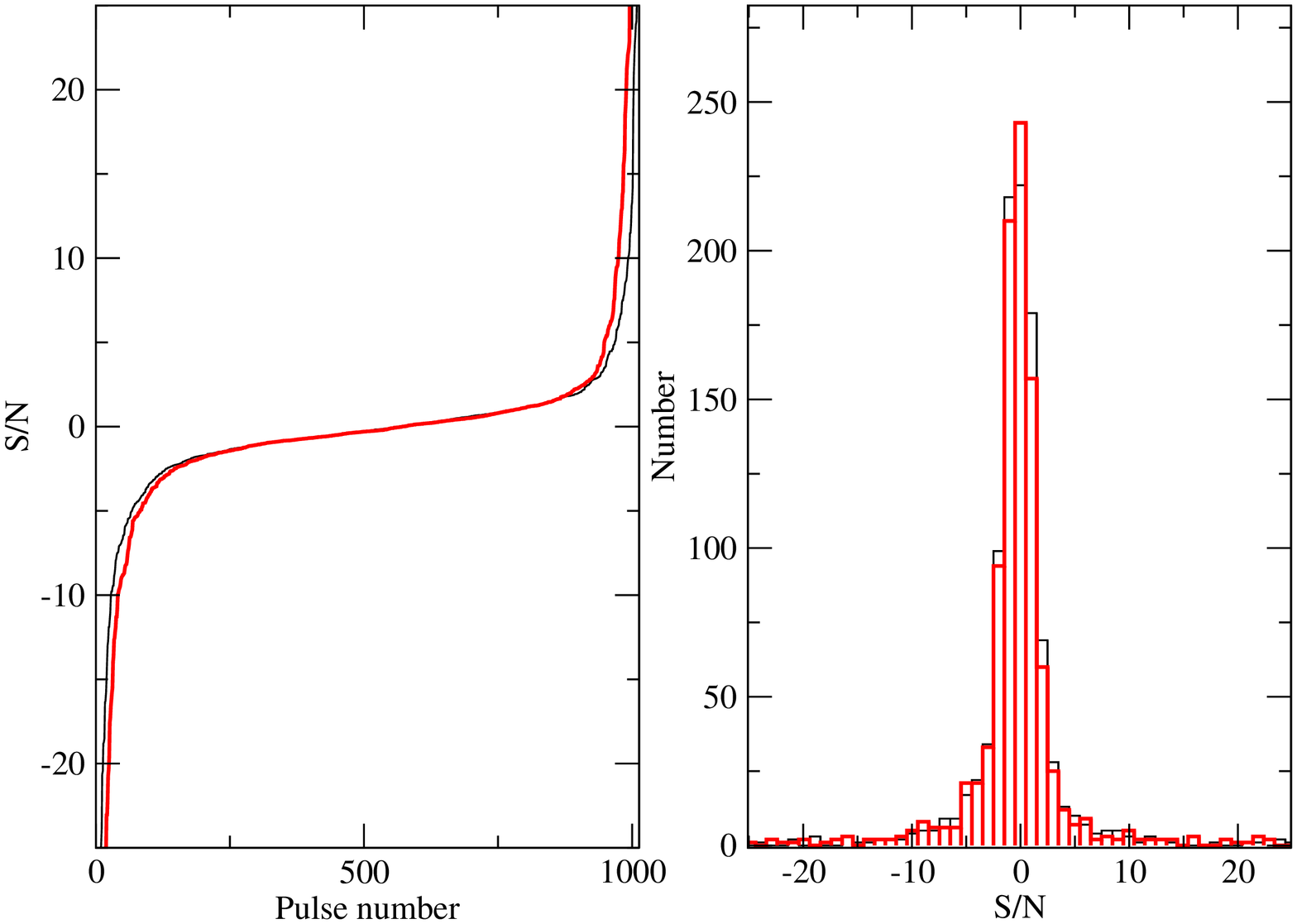}} }
\caption{The cumulative (left) and normal (right) histogram of the
  real (thin black line) and simulated (thick gray line) circular
  polarization data from pulsar B1133+16.}
\label{1133}
\end{figure*}

In observations of single pulses, four Stokes parameters ($I$, $Q$,
$U$, $V$) are recorded per phase bin of the pulse period. The
off-pulse data have a distribution which is Gaussian with a standard
deviation, $\sigma_n$, determined by the system noise and are
manipulated to ensure a mean of zero.  For any symmetrical
distribution of $V$ with mean $V_o$, the integrated signal after $N$
pulses will be $N\times V_o$, the rms of the off-pulse noise will be
$\sqrt{N}\times\sigma_n$ and hence the resultant signal to noise ratio
(S/N) will be $\sqrt{N}\times V_o/\sigma_n$.

To form the integrated profile of $|V|$ we sum the magnitude of $V$ in
the appropriate phase bin for each single pulse.  Considering the case
where a given phase bin has a Gaussian distribution in $V$ with mean
$V_o$ and standard deviation $\sigma_o$, one can show that $|V|$ will
have a mean $\overline{|V|}$ given by
\begin{equation}
\overline{|V|}=\sqrt{\frac{2}{\pi}}\, \sigma e^{\left(\frac{-V_o}{\sqrt{2}
 \sigma}\right)^2}+\frac{V_o}{\sqrt{\pi}}\int_{\frac{-V_o}{\sqrt{2}
 \sigma}}^{\frac{V_o}{\sqrt{2} \sigma}}e^{\left(\frac{-(V-V_o)}{\sqrt{2}
 \sigma}\right)^2}dV,
\end{equation}
where $\sigma=\sqrt{\sigma_o^2+\sigma_n^2}$.
This can be re-written as
\begin{eqnarray}
\overline{|V|} =\sqrt{\frac{2}{\pi}}\, \sigma
 e^{\left(\frac{-V_o}{\sqrt{2} \sigma}\right)^2}+V_o
 {\rm erf}\left(\frac{V_o}{\sqrt{2} \sigma}\right),
\end{eqnarray}
with $\rm erf$ being the error function, defined as:
\begin{equation}
{\rm erf}(x)=\frac{2}{\sqrt{\pi}}\int_{0}^{x}e^{-t^2}dt.
\end{equation}
Consider equation 1 for the off-pulse emission which has $V_o=0$ and
$\sigma=\sigma_n$. In this case the mean off-pulse signal
which we denote $\overline{|V|}_n$ is simply
\begin{equation}
\overline{|V|}_n = \sqrt{\frac{2}{\pi}}\, \sigma_n.
\end{equation}
In order that the off-pulse region of the profile of $|V|$ have a zero
mean therefore, we must subtract $\overline{|V|}_n$ from each phase bin of
each profile.  The S/N in each bin of the $|V|$ profile is then
\begin{equation}
{\rm S/N} = \left(\frac{\overline{|V|}}{\sigma_n} -
\sqrt{\frac{2}{\pi}}\right) \sqrt{N}
\end{equation}

When we refer to $|V|$ hereafter, we are actually referring to
$\overline{|V|}$ from the above equations and, in particular, when we
refer to the S/N of $|V|$, it is given by Equation 5. Figure \ref{snr}
shows the S/N in $|V|$ compared to $V$ as a function of $V_o$ in the
single pulses, assuming $\sigma_n=1$ and $\sigma_o=0$ (i.e. that the
intrinsic $V$ distribution is a delta function).  It is apparent that
for values of $V_o$ greater than $\sim 1$, the S/N in $|V|$ builds up
linearly but for values below this the S/N in $|V|$ builds up at a
slower rate.  However, because we are forced to subtract the baseline,
the signal to noise in $|V|$ is always less than in $V$ (when
$\sigma_o=0$).  Figure \ref{sigma} shows the S/N in $|V|$ as a
function of $\sigma_o$ for four values of $V_o$. The S/N in $|V|$
increases with $\sigma_o$,whereas, because the distribution is
symmetrical, the S/N in $V$ is independent of $\sigma_o$. There are
two main points to be gleaned from these figures.  The first is that
the S/N in $|V|$ cannot be less than the case where the distribution
of $V$ is a delta function (i.e. $\sigma_o=0$). We can therefore
construct a `minimum profile' of $|V|$ from the mean $V$ in each phase
bin. The second point is that, under the assumption of a Gaussian
distribution of $V$, one can determine $\sigma_o$ by comparing the S/N
in $|V|$ and $V$.

As discussed in the previous section, the distribution of $V$ may be
log-normal rather than Gaussian.  In this case, the convolution of the
log-normal distribution with the Gaussian noise is not trivial
analytically. We have therefore treated the problem numerically and
present the results in the right hand panel of Figure \ref{sigma}. The
median S/N is the same as the mean S/N in the left hand panel of the
figure, and the sigma is now in the log.  It can be seen that the mean
$|V|$ rapidly increases as a function of $\sigma_o$, much more rapidly
than in the Gaussian case. However, the mean of $V$ also increases
rapidly with $\sigma$. As a result, for typical values of $\sigma$
($\sigma <1$), it remains that $|V|<V$.

So far, we have only considered unimodal distributions of $V$. In
Karastergiou et al. (2003a) and KJLME it was shown that large values
of $V$ of either handedness occur irrespective of the dominant OPM,
which suggests that the handedness of $V$ may be set randomly. We
therefore make the simplistic assumption that $V$ is always generated
with a positive sign and introduce the parameter $F$ to denote the
fraction of data in which the sign of $V$ changes
(i.e. $0.0<F<1.$). It is for distributions with $F\approx 0.5$ that
$|V|$ becomes a very powerful tool as the S/N in $|V|$ is independent
of $F$ whereas the S/N in $V$ is reduced by a factor $1-2F$.

Figure \ref{distributions} shows examples of simulated data chosen
from bi-modal Gaussian and log-normal distributions, where
$F=0.25$. In cases where the S/N is high (left-hand side panels), the
two components are clearly separated. However, in cases of low S/N
(right-hand side panels), the convolution with the noise merges the
distributions. This means that in the Gaussian, low S/N case (top
right), the deconvolution to determine the underlying distributions
does not lead to a unique solution. In the log-normal, low S/N case
(bottom right) however, the parameters of the parent distribution are
much better constrained: it must be bimodal ($F\neq1$ and $F\neq0$)
and $V_o$ and $\sigma$ are constrained by the characteristics of the
positive and negative tails of the distribution. 
We note that in real pulsar observations, distributions like the high
S/N cases on the left hand side have not been observed. However, this
may be purely due to the low S/N in single-pulse observations. An
increase in sensitivity in the future will reveal whether such
``separated'' distributions are real.

What remains to be tested is whether the three parameters we
introduced can describe observational data.  Each set of $V_o$,
$\sigma$ and $F$ values are used to simulate a particular
distribution.  We then apply a Kolmogorov-Smirnoff (K-S) test (Press
et al. 1992\nocite{ptvf92}) to compare the real and simulated
distributions of $V$.  The K-S test returns the probability $P$ that
the two compared datasets do not originate from the same
distribution. A concrete example involves PSR B1133+16 data at 4.85
GHz. In the relevant profile of Figure 1 of KJMLE, $|V|$ has a maximum
in the leading component and we examine this particular phase bin.
The S/N in the integrated profile (over a total of 1013 pulses) of
that bin is $V=-3.8$ and $|V|=82.1$. We find that the intrinsic $V$
distribution which best reproduces the data is a log-normal
distribution with $|V_o|=0.43, \sigma= 0.95$ and $F=0.64$. Figure
\ref{1133} shows how good the simulation is in fitting the observed
data. The thick line in the two histograms corresponds to the real
data and the thin line to the simulation.
The claim that the underlying $V$ distribution is bimodal is somewhat
contrary to the visual impression given by Figure \ref{1133}. However,
in the context of our model where the distribution of $|V|$ is either
Gaussian or log-normal and a sign change occurs in a fraction of the
pulses, this particular distribution can only be adequately described
by a bimodal solution. This is no surprise, given that the observed
$V$ distribution closely resembles the bottom right simulated
distribution of Figure \ref{distributions}. The fact that this
solution describes the observed data very well, shows that our model
works.
Also, the K-S test is more sensitive around the median value which is
the reason our method overestimates $\sigma$, causing the disagreement
seen in the tails of the cumulative distribution.

From the above description we can see how the combination of $|V|$ and
$V$ can reveal details of the underlying distributions in each phase
bin, without needing to perform the tedious task of constructing and
analysing the $V$ distribution itself.  In summary, there are 3
possibilities which can occur.  First, that $|V|$ is significantly
larger than the magnitude of $V$.  This can occur either because the
parameter $F$ is close to 0.5 and/or because the width of the
underlying distribution is large.  The second possibility is that
$|V|$ and $V$ are about the same magnitude. In this case, $F$ must be
near 0 or 1 and/or the width of the underlying distribution must be
small relative to the mean.  Finally, $|V|$ can be less than the
magnitude of $V$. This can occur in low signal-to-noise regimes (cf
Figure \ref{snr}) provided that the width of the distribution is not
large.

\section{Enhanced pulsar profiles}

Having shown that our simple model describes the data well, we show 13
profiles of pulsars including $|V|$ and extract conclusions about the
behaviour of circular polarization in single pulses. 

The data consist of observations of single pulses from 13 northern
pulsars, 5 at 1.41 GHz and 8 at 4.85 GHz, to complement the sample of
KJMLE. We used the Effelsberg 100-m telescope in observations that
took place between July and September 2002. The data were carefully
calibrated according to a revised scheme based on von Hoensbroech \&
Xilouris (1997\nocite{hx97}). Both the 4.85 and the 1.41 GHz system
had a system equivalent flux density of $\approx20$ Jy. We used a
bandwidth of 500 MHz at 4.85 GHz and a choice between 10, 20 and 40
MHz at 1.41 GHz. The pulsars in our sample were chosen to have a
reasonably small dispersion measure (DM) due to the limitations of the
backend used and the choice of bandwidth at 1.41 GHz was based on each
particular DM.

In Figures 5 and 6, we plot the average profile of the total power, of
circular polarization $V$ and the profile of $|V|$, which is computed
as described in Section 3, by subtracting the noise offset from each
single pulse. For clarity, we have not included the linear
polarization nor the position angle swing for these pulsars, but we
have used all the polarization information contained in the integrated
profiles and the single pulses where relevant.

\begin{figure*}
\begin{center}
\begin{tabular}{cc}
\resizebox{0.5\hsize}{!}{\includegraphics{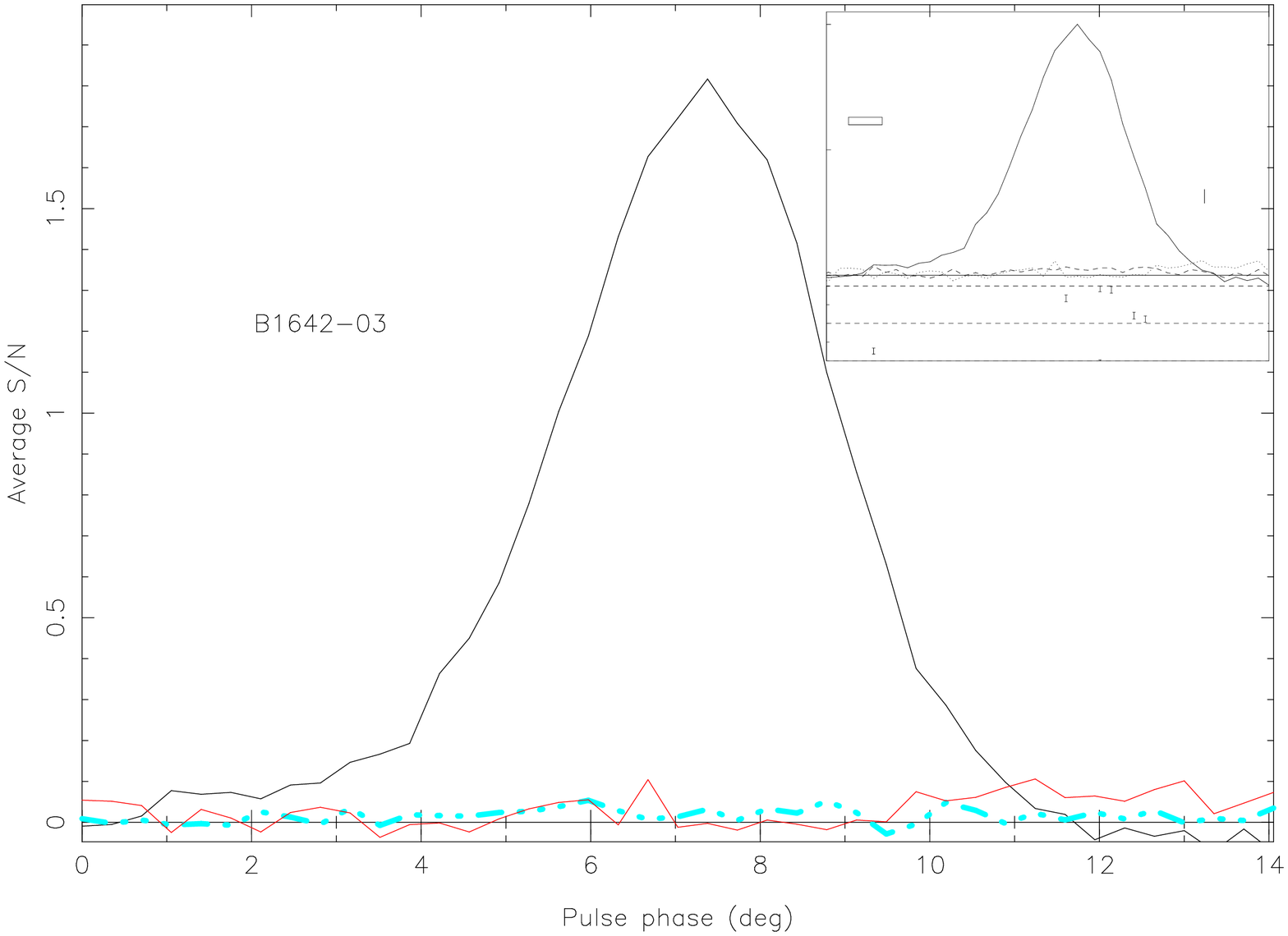}} &
  \resizebox{0.5\hsize}{!}{\includegraphics{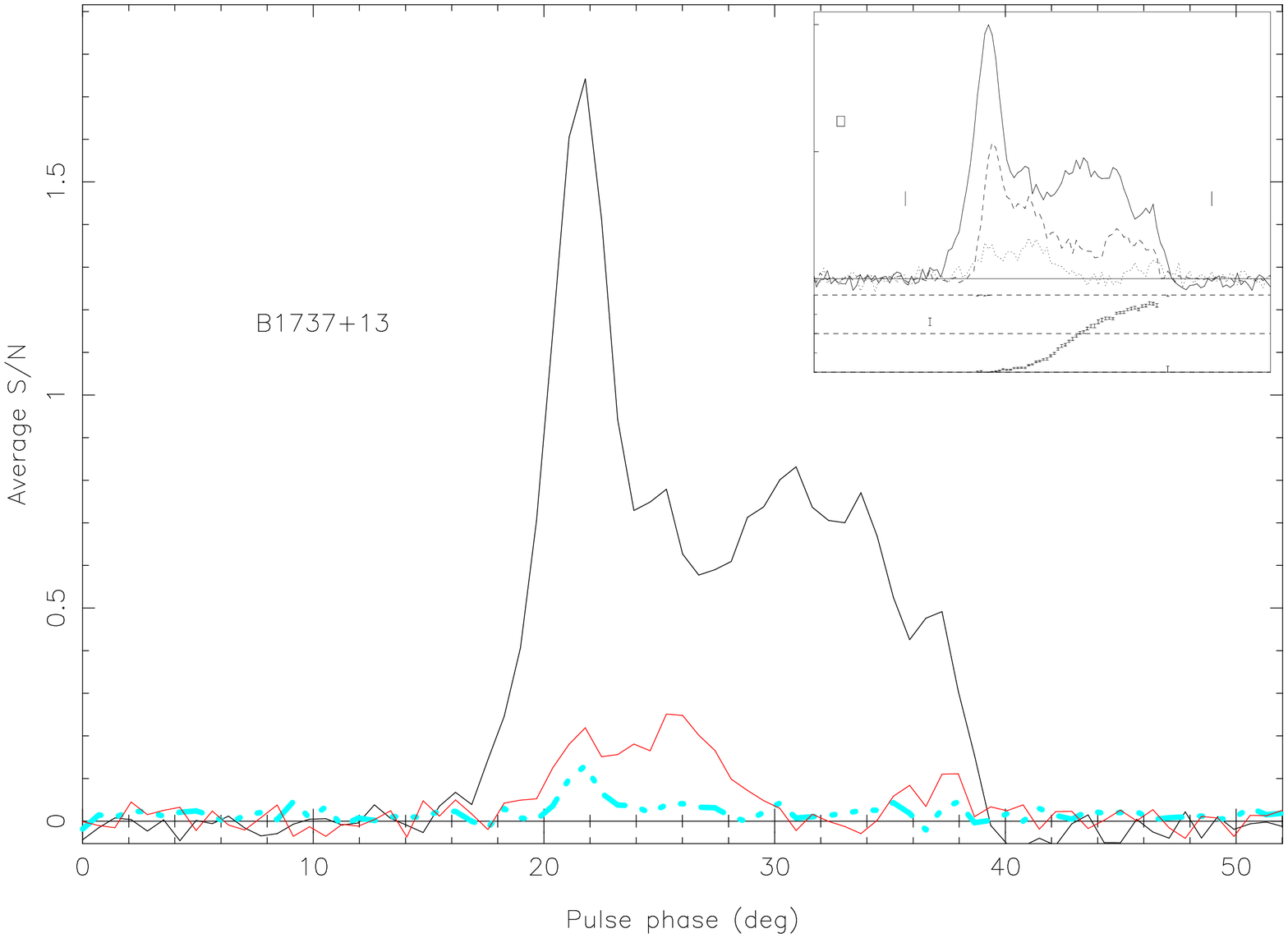}}\\
  \resizebox{0.5\hsize}{!}{\includegraphics{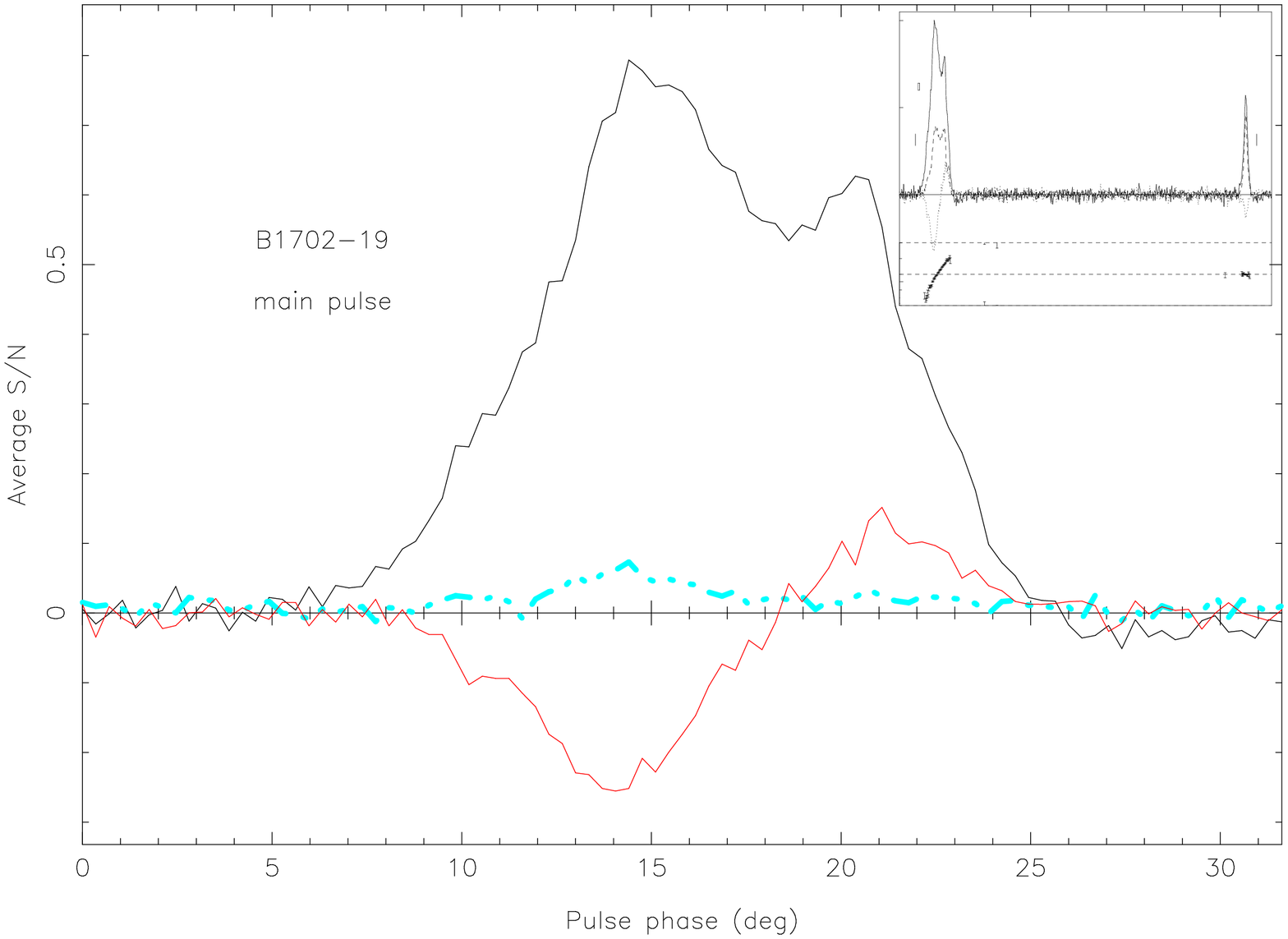}} &
  \resizebox{0.5\hsize}{!}{\includegraphics{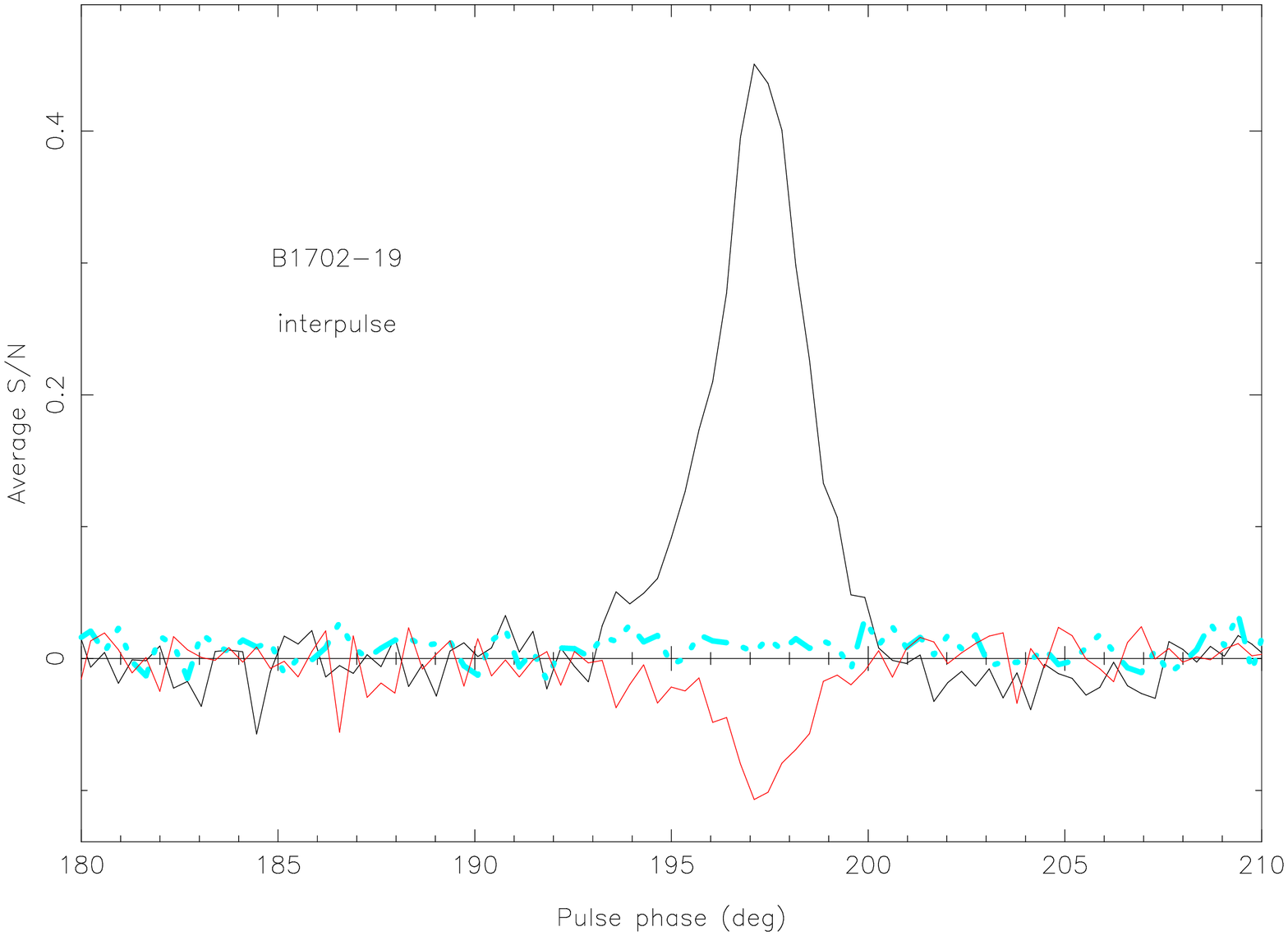}}\\
  \resizebox{0.5\hsize}{!}{\includegraphics{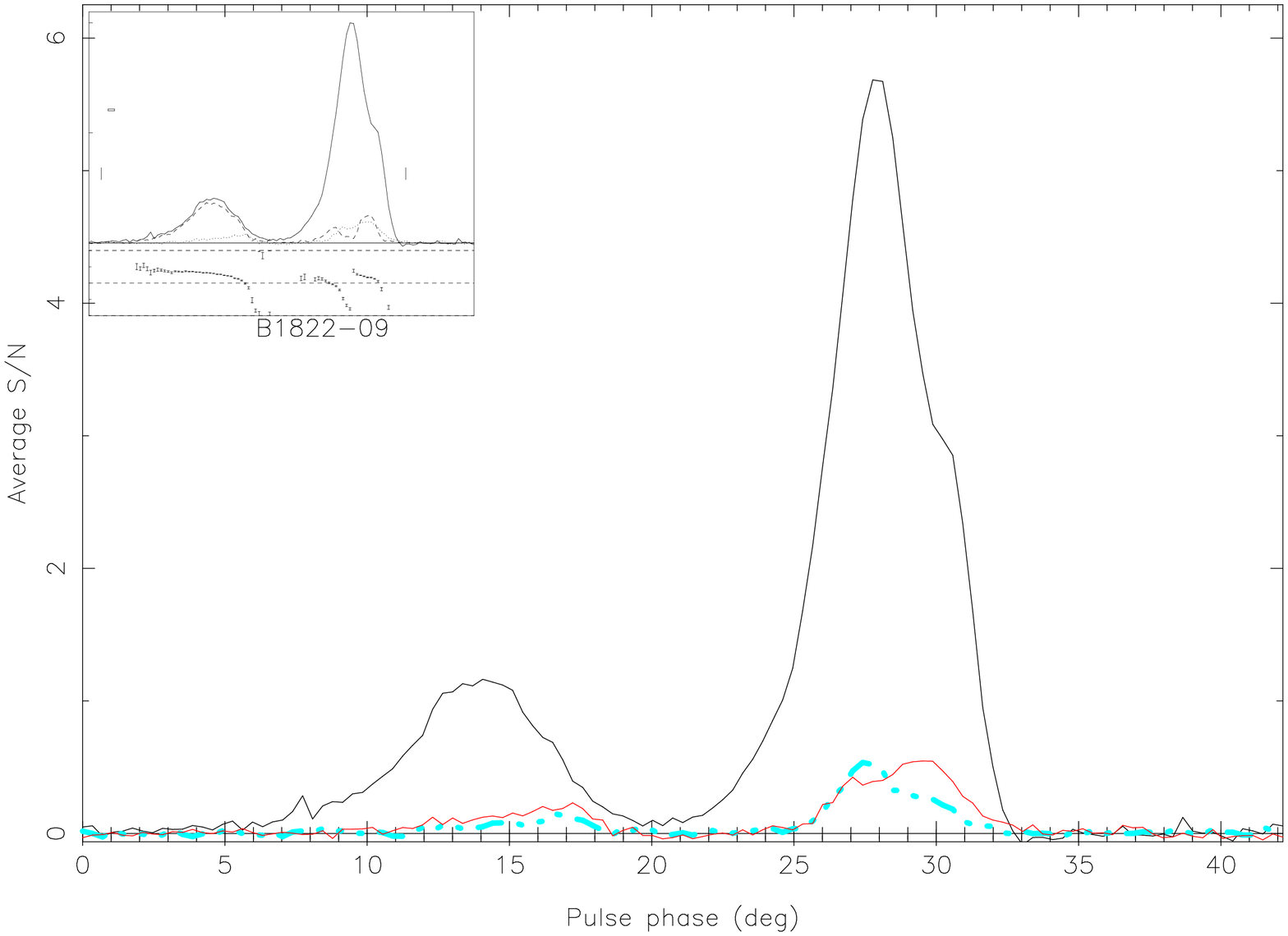}} &
  \resizebox{0.5\hsize}{!}{\includegraphics{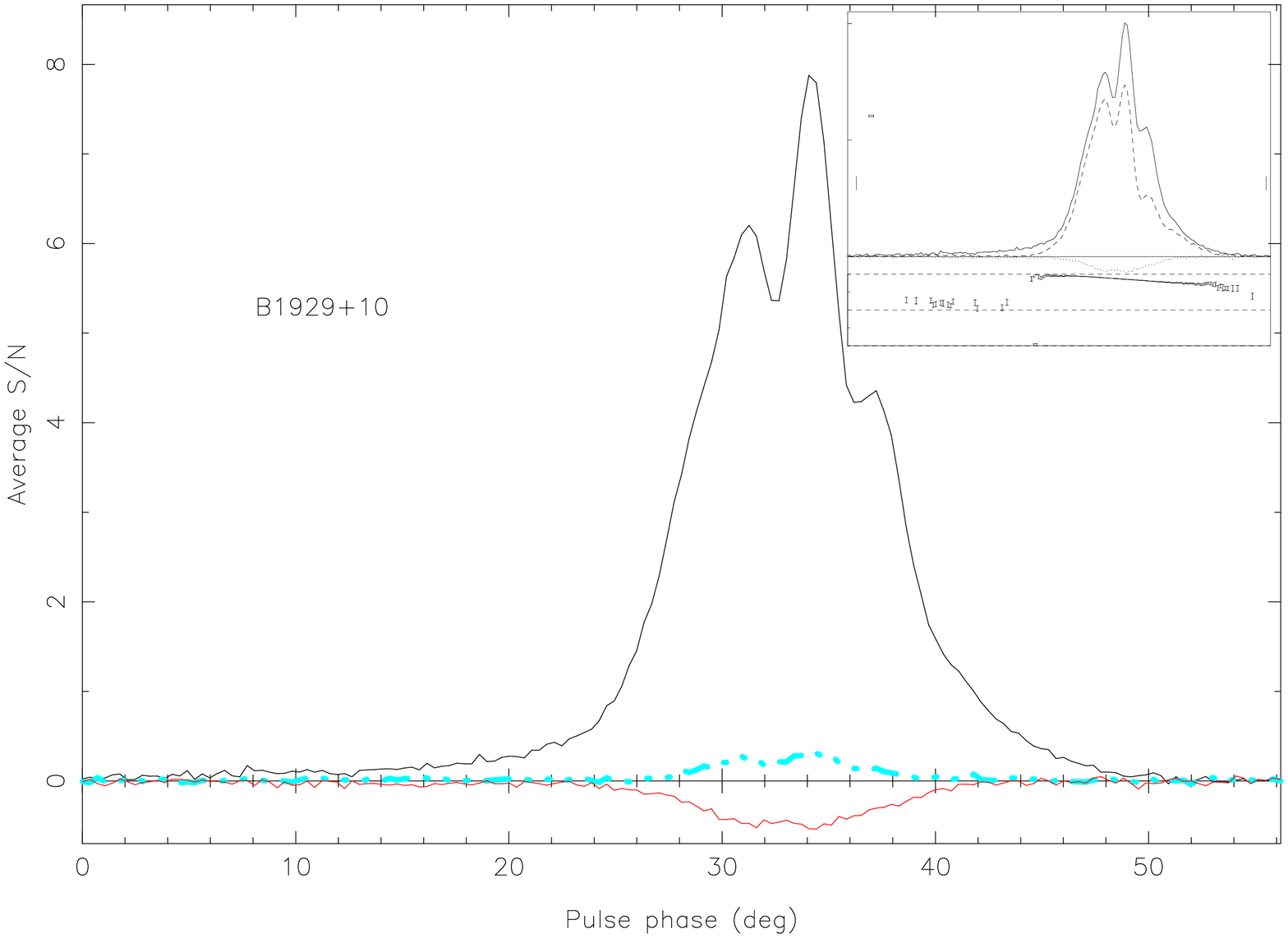}}\\
\end{tabular}
\end{center}
\caption{Integrated profiles of 5 pulsars at 1.41 GHz. Both the main
  pulse and interpulse are shown for PSR B1702$-$19. The average S/N
  of the total intensity $I$ and the circular polarization $V$ are
  shown with thin lines, the $|V|$ profile is shown with a thicker,
  dot-dashed line. The inserts represent the full polarization
  profiles, including $L$ (dashed line), and the PA ranging from
  $-90^o$ to $90^o$.}
\end{figure*}
\subsection{Profiles at 1.4 GHz}
{\bf PSR B1642$-$03}: The integrated profile of PSR B1642$-$03 at 1.4
GHz consists of a single component which shows virtually no linear or
circular polarization.  Lyne \& Manchester (1988, hereafter
LM\nocite{lm88}) classify this pulsar as a core single.  We find that
the profile of $|V|$ is just like that of $V$, with $S/N\approx 0$
across the entire component and, furthermore, we do not see any highly
linearly polarized single pulses. We conclude that the individual
pulses are not highly circularly polarized. The evidence points to the
fact that the two competing orthogonal modes are equally strong and
highly correlated.

{\bf PSR B1702$-$19}: The average profile of PSR B1702$-$19 shows a
swing of $V$ from negative to positive.  The profile of this pulsar is
classified as cone-dominated in LM. However, the steep position angle
swing of the linear polarization and the presence of a phase bin near
the centre where both $V$ and $|V|$ are zero make it more likely that
this is a core component.  At the pulse phase where $V$ changes
handedness, $|V|\approx 0$. This is a common feature pointed out by
KJMLE, implying that at this particular phase bin the circular
polarization is consistently a small fraction of $I$.  The S/N of the
single pulses is low and we cannot draw many conclusions from the
profile of $|V|$. However, it is likely that the swing of $V$ is a
feature of the single pulses. The interpulse, shown in the next
panel of the same figure, is almost 100\% linearly polarized and
only weakly circularly polarized. The low S/N in $V$ forces the
profile of $|V|$ to be zero, so the conclusion is that there are no
pulses where the interpulse is highly circularly polarized.

{\bf PSR B1737+13}: There are a number of pulsars with profiles that
consist of both cone and core components. In PSR B1737+13, the profile
also exhibits a very smooth position angle swing and a relatively high
degree of linear polarization, although again the circular
polarization across the profile is low. More specifically, in the
leading component, $V$ has a peak value of 0.22. If the intrinsic $V$
distribution were a delta-function, Figure 1 suggests that $|V|$
should have a very small value ($<<0.1$). In reality, the measured
value is $|V|\approx 0.15$, which would be the result of a $V$ delta-function
at $V_o\approx 0.5$. We therefore can deduce that either a sign-changing
effect reduces $V$, or the distribution of $V$ has some intrinsic
width that makes the distribution wider and gives rise to some values
of opposite handedness. The linear polarization under the peak is
$\approx 50\%$ and makes for a good comparison with the next
component. In that, $L$ is a significantly higher fraction of $I$, but
$|V|$ is very weak compared to $V$, implying a narrow distribution
around $V_o$. In the centre of the profile of this pulsar also,
$V=|V|=0$.

{\bf PSR B1822$-$09}: The profile of this pulsar consists of the main
pulse and an interpulse, which both exhibit a peculiar moding
behaviour (Fowler \& Wright 1982 \nocite{fw82}). We focus only on the
main pulse, mainly due to the low S/N of the interpulse. LM suggest
that both the leading and trailing component of the main pulse are
conal. The leading component is highly linearly polarized.  As far as
the circular polarization is concerned, the leading component of the
main pulse is weak in both $V$ and $|V|$ and it is likely that the
circular polarization is generally low in the single pulses. In the
trailing component of the main pulse, however, $|V|$ is as high as
$V$.  We also see evidence for a bi-modal position angle distribution.
In this component therefore, it appears as if $V$ is often large and
of variable sign in order to give rise to a large $|V|$.

{\bf PSR B1929+10}: The profile of PSR B1929+10 is almost entirely
linearly polarized. The swing of position angle across the pulse is
rather flat and the profile is classified as conal by LM. Both $V$ and
$|V|$ in the integrated profile are low. The results for this pulsar
are very similar to that seen in the Vela pulsar (Kramer et
al. 2002). A single orthogonal mode dominates throughout, the single
pulses are highly linearly polarized and have a largely constant sign
of circular polarization.

\begin{figure*}
\begin{center}
\begin{tabular}{cc}
\resizebox{0.5\hsize}{!}{\includegraphics{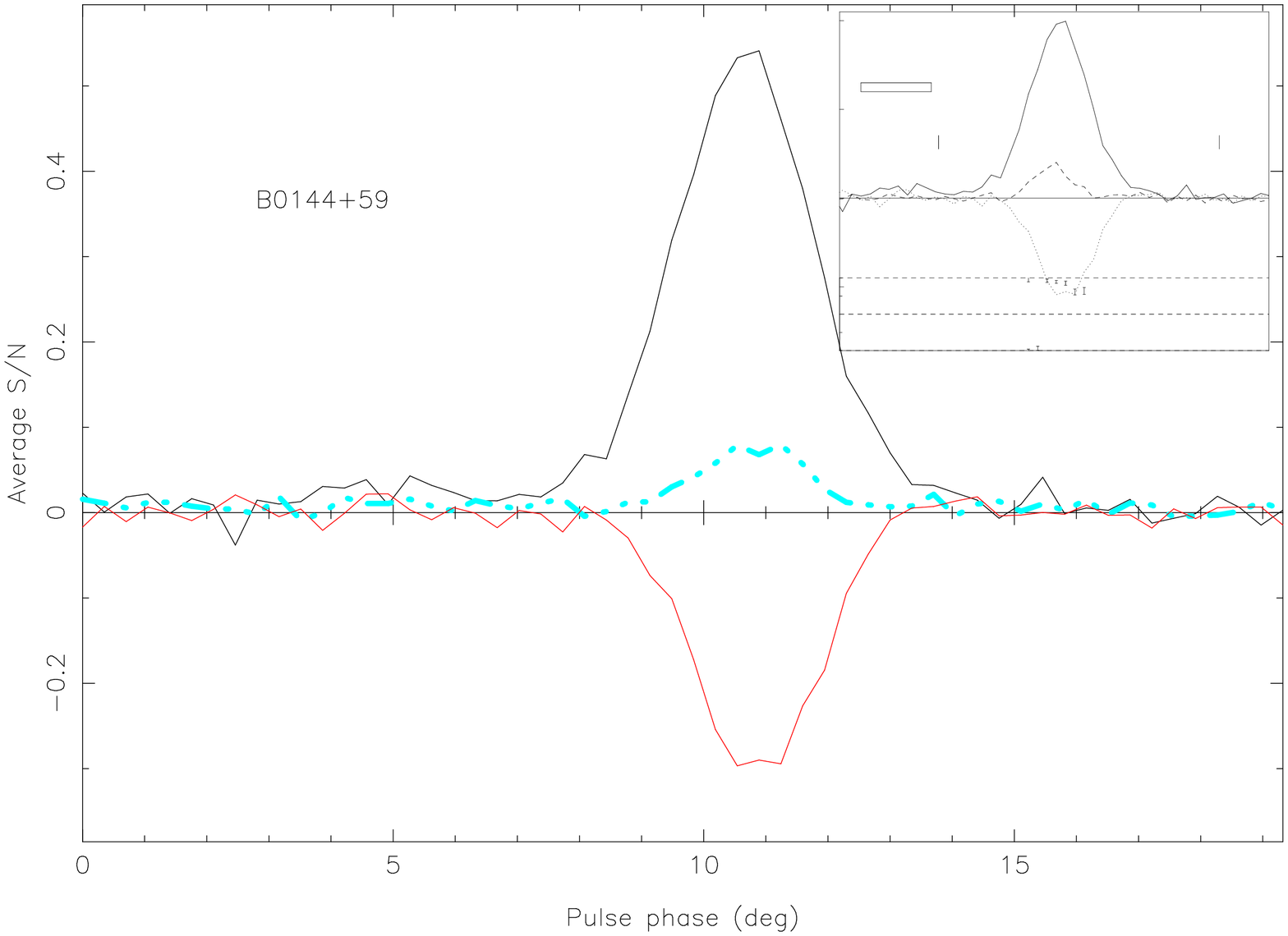}}
  &
  \resizebox{0.5\hsize}{!}{\includegraphics{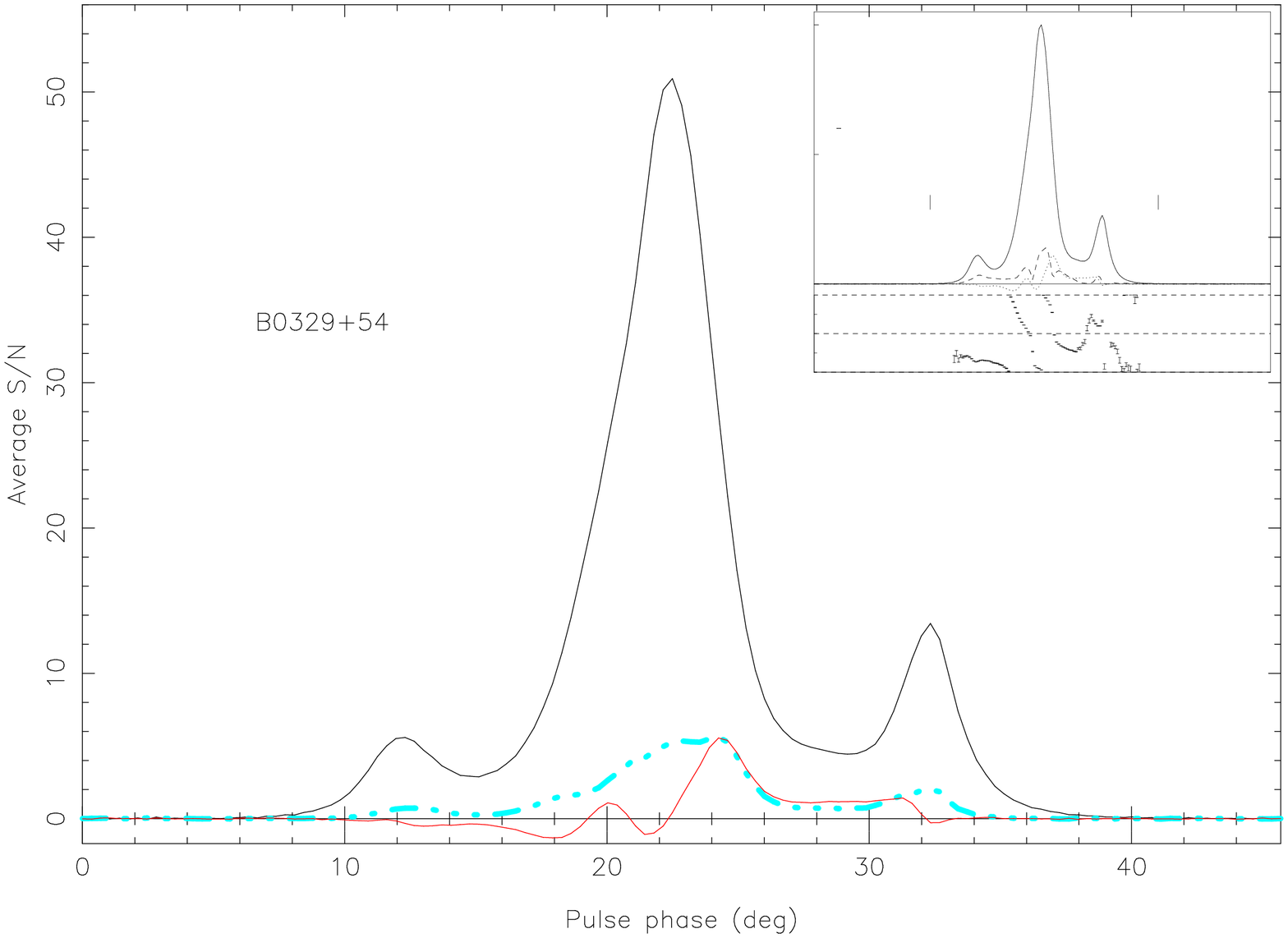}}\\
  \resizebox{0.5\hsize}{!}{\includegraphics{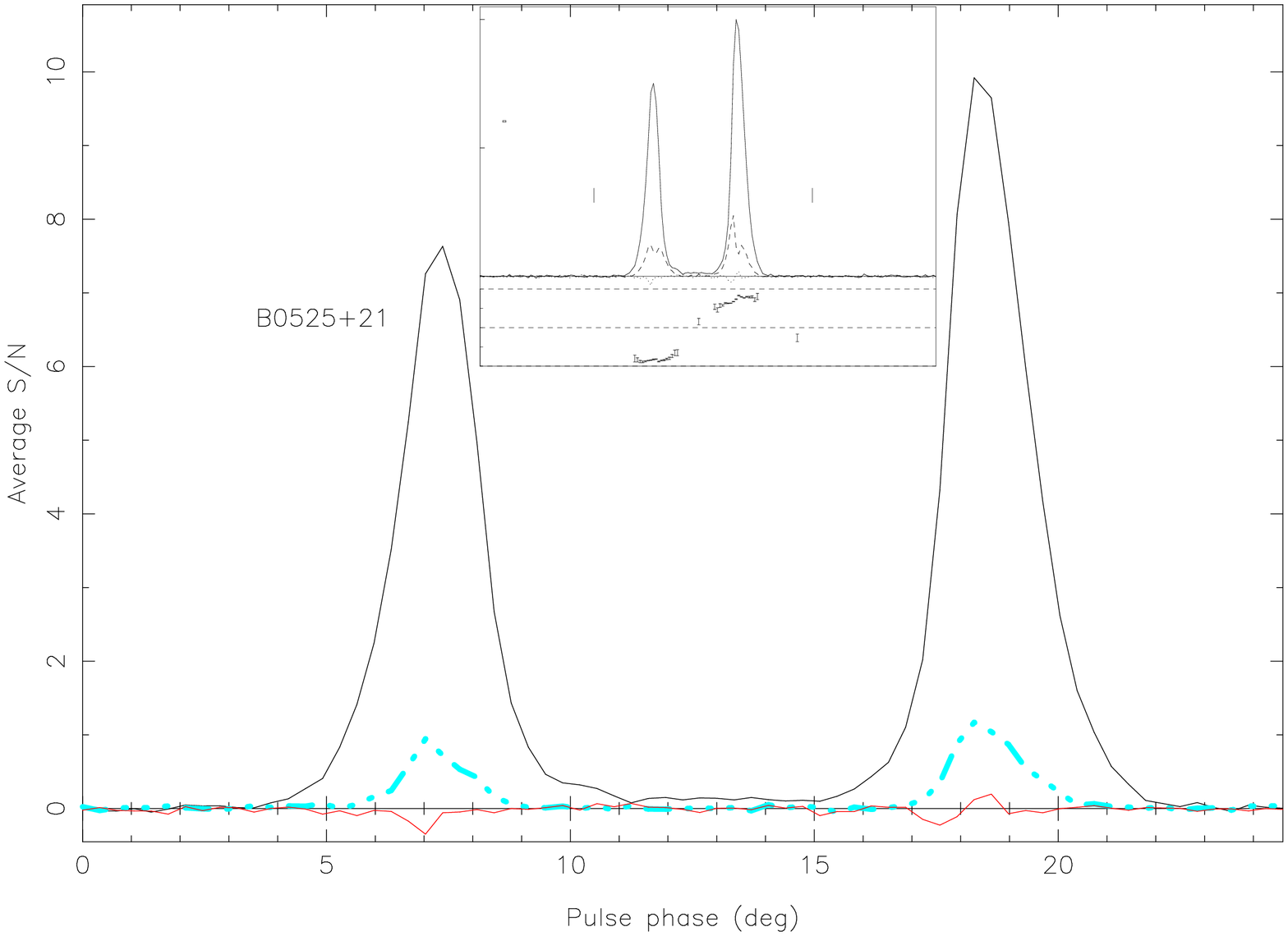}}
  &
  \resizebox{0.5\hsize}{!}{\includegraphics{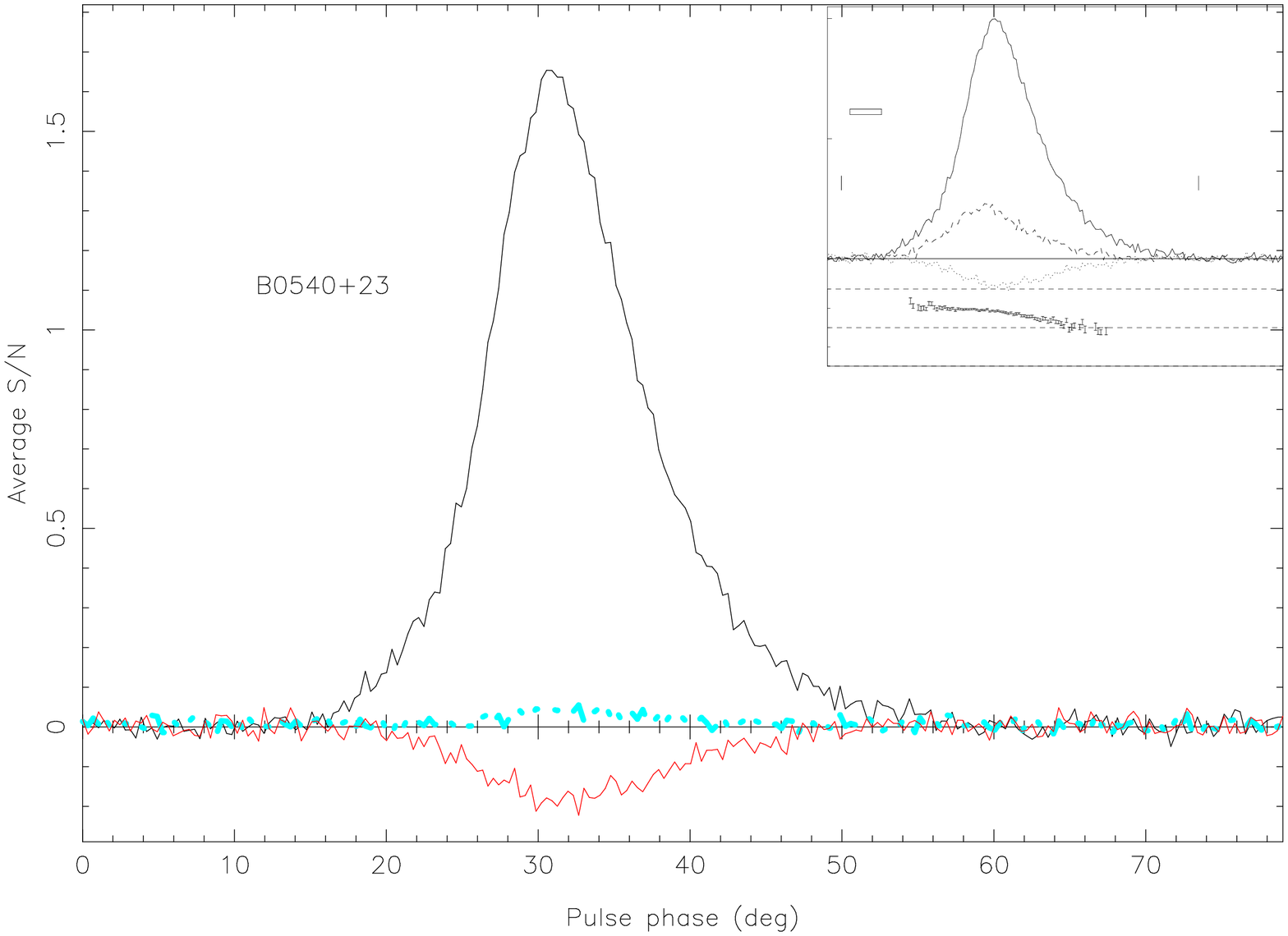}}\\
  \resizebox{0.5\hsize}{!}{\includegraphics{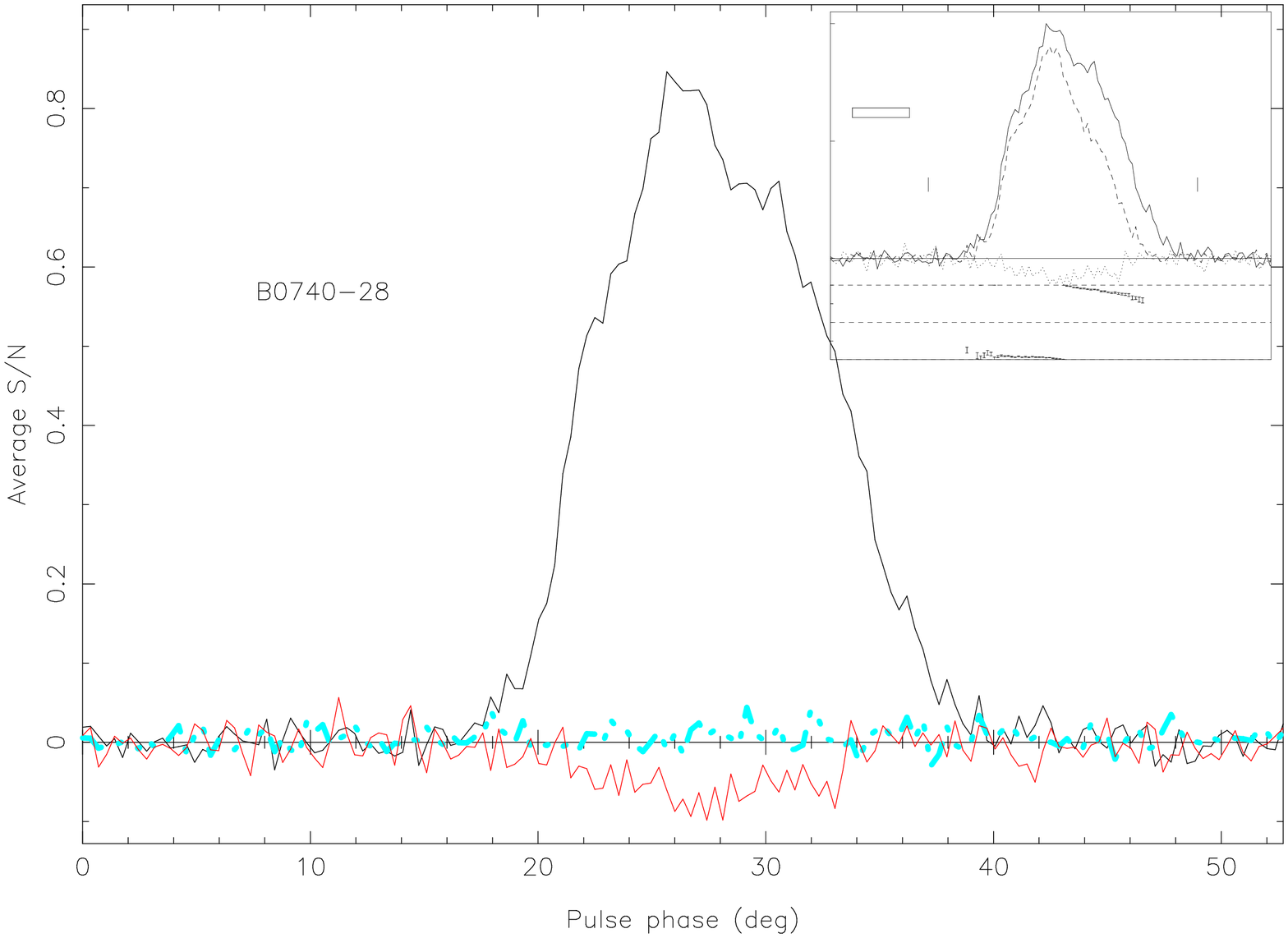}}
  &
  \resizebox{0.5\hsize}{!}{\includegraphics{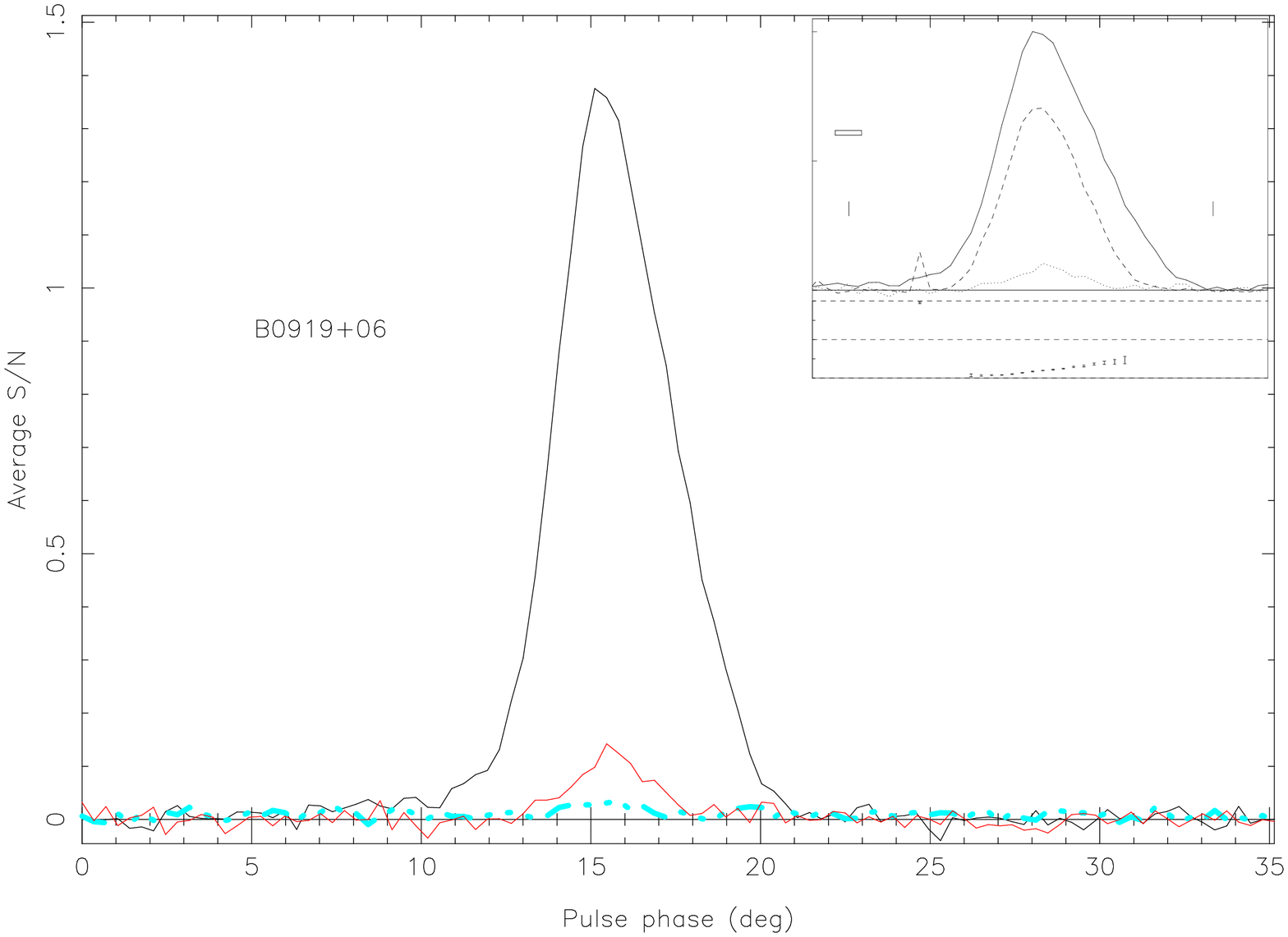}}\\
\end{tabular}
\end{center}
\caption{Integrated profiles of 6 pulsars at 4.85 GHz. The lines and
  the inserts are explained in Figure 5.}
\end{figure*}
\addtocounter{figure}{-1}
\begin{figure*}
\begin{center}
\begin{tabular}{cc}
\resizebox{0.5\hsize}{!}{\includegraphics{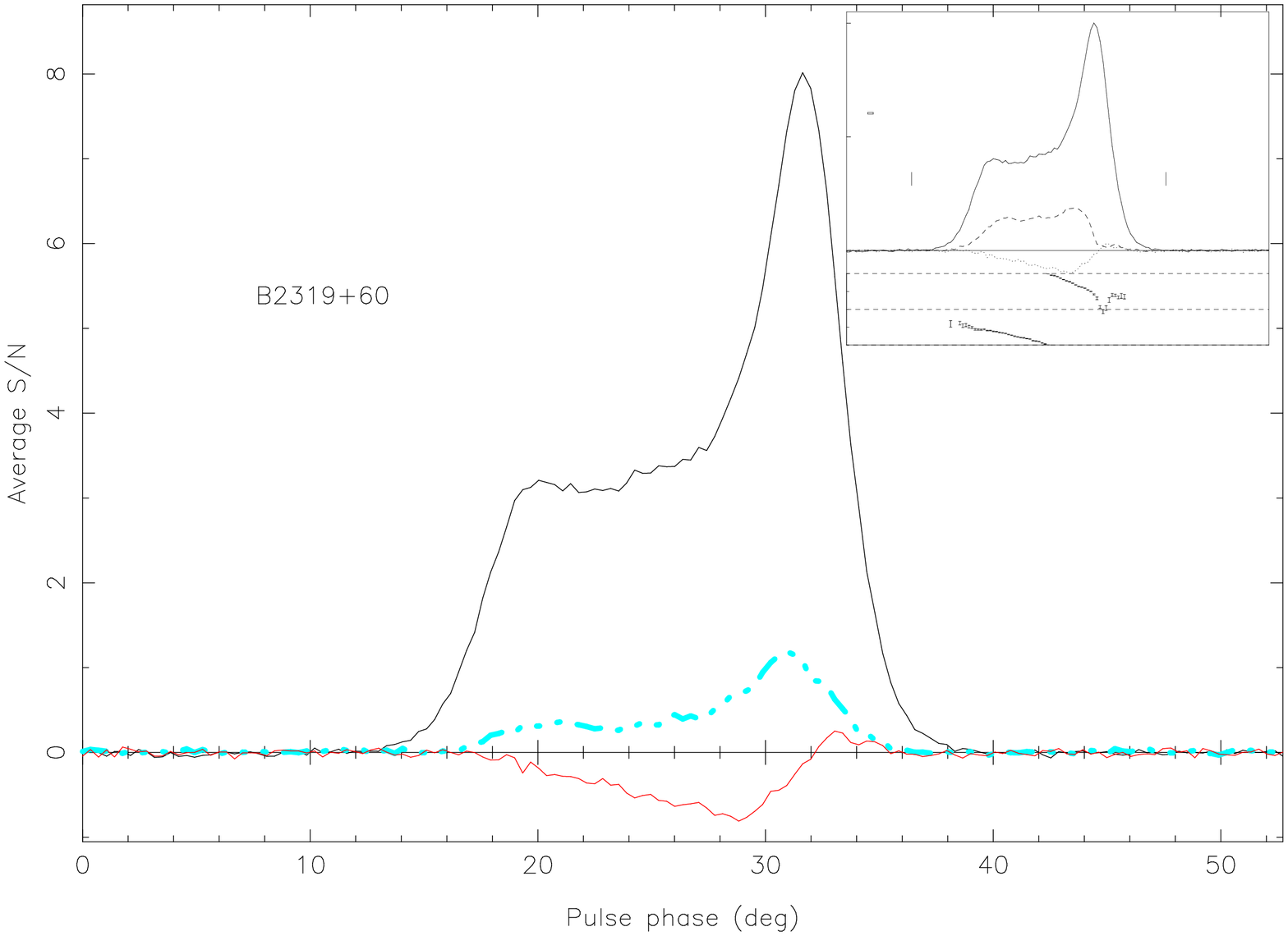}}
  &
  \resizebox{0.5\hsize}{!}{\includegraphics{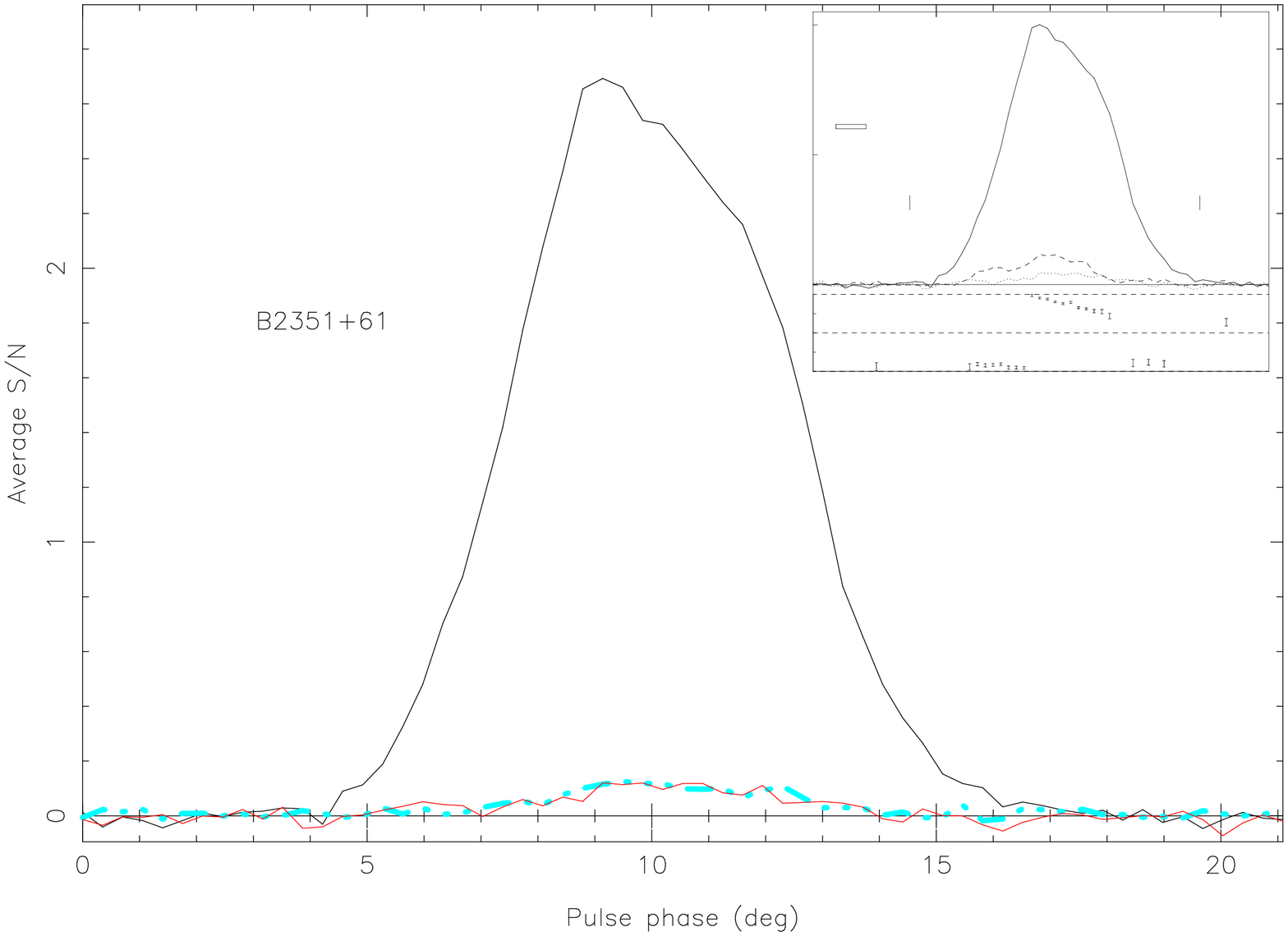}}\\
\end{tabular}
\end{center}
\caption{...continued.}
\end{figure*}
\subsection{Profiles at 4.85 GHz}
{\bf PSR B0144+59}: PSR B0144+59 has not been classified by LM, but
the flat position angle swing of the linear polarization is
characteristic of conal emission.  The integrated profile has a very
high degree of right-hand circular polarization, one of the highest
fractional circular polarizations of any pulsar.  The $|V|$ profile is
that expected from the low S/N and an almost constant value of $V$. It
seems unlikely that there are single pulses with strong positive
circular polarization.

{\bf PSR B0329+54}: The polarization profile of PSR B0329+54 at 4.85
GHz is very complex. The linear polarization drops suddenly to local
minima at two pulse phases, the handedness of the circular
polarization changes a number of times and the position angle shows a
number of kinks and jumps (as has been noted before by other authors,
e.g. Gil \& Lyne 1995\nocite{gl95}). The profile of $|V|$ is much less
complicated than the profile of $V$. In fact, it traces the average
profile of the total power well, so that the quantity $|V|/I$ remains
fairly constant across the pulse. In the leading and trailing
components, $|V|>V$ and at the same time, competing OPMs are evident
through the reduced linear polarization and OPM jumps in the single
pulses. Also, in the leading part of the middle component, $V$ changes
handedness on three occasions, OPM jumps and local minima of $L$ are
seen and $|V|$ is significantly greater in amplitude than $V$. In
fact, at the phase bin just after $22^o$ in Figure 6, the average S/N
in $|V|$ is $\approx 6$ whereas $V_o=0$. The actual distribution
resembles the PSR 1133+16 example of the previous section: it is
clearly log-normal and bimodal, while the linear polarization has a
local minimum. The same thing holds for the phase bin at $32^o$.

{\bf PSR B0525+21}: PSR B0525+21 is a long period pulsar with a
distinctive double profile of conal components. The overall
polarization in the integrated profile is low, neither component shows
significant $V$.  However, the profile of $|V|$ has a reasonably large
amplitude.  This implies that there must be a number of highly
circularly polarized single pulses, of either handedness. At the same
time, even though no orthogonal jumps are seen in the position angle
in the integrated profile, the position angle has a bimodal
distribution in the single pulses.

{\bf PSR B0540+23}: LM classified the profile of PSR B0540+23 as conal
based on the flat position angle swing. There are no OPM jumps in the
position angle and there is a moderate amount of linear polarization
in the profile.  Despite the fact that at 4.85 GHz, it shows a
moderate amount of circular polarization, the profile of $|V|$
indicates that there generally is a constant sign of $V$ in the single
pulses.  The results for this pulsar therefore resemble those for Vela
and PSR B1929+10.

{\bf PSR B0740$-$28}: LM classify this pulsar as having conal
emission. There is a high degree of polarization and no evidence of
OPM jumps in the single pulses. Similar to PSR B0540+23, this pulsar
is very weak in both $V$ and $|V|$.  Although the S/N is not high in
total intensity in this pulsar, it can be compared directly with PSR
B1702$-$19 which has a similar S/N but has a significant amount of
$|V|$. The profile of $|V|$ suggests that the circular polarization in
the single pulses is generally low with few high values of either
handedness seen.

{\bf PSR B0919+06}: PSR B0919+06 is considered to have partial cone
emission by LM, has a very similar profile to PSR B1929+10 and similar
polarization features to Vela. The profile of $|V|$ suggests that the
single pulses are only moderately circularly polarized.

{\bf PSR B2319+60}: PSR B2319+60 is a strong pulsar with both core and
cone emission, according to LM.  At first glance the profiles of $V$
and $|V|$ at 4.85 GHz look like mirror images of each other. A closer
look, however, at the trailing edge of the profile reveals that $|V|$
remains large at the phase where $V$ changes handedness. This change
in handedness in $V$ coincides with a kink in the position angle
swing, which is not quite an orthogonal jump.  At the same phase, the
distribution of position angles in the single pulses is bimodal. As
with PSR B0525+21, there are many single pulses with significant $V$
of either handedness.

{\bf PSR B2351+61}: PSR B2351+61 has low linear polarization and also
shows an OPM jump in the position angle swing. As with PSR B1133+16,
the lack of polarization at high frequencies is indicative of a lack
of correlation between the mode strengths. The single pulses show a
bimodal distribution of position angles throughout the pulse profile.
The profile of $|V|$ is slightly greater than that of $V$.  This shows
that there are pulses with significant $V$ of either sign in the
single pulses. This is similar to PSRs B1133+16 and B0525+21.

\begin{table*}
\begin{tabular}{ccccccc}
\hline PSR name & \multicolumn{2}{c}{Classification} & Degree of &
Bimodal PA & $|V|=V=0$ & $|V|>V$\\ 
& cone & core & polarization & distributions & in core &\\ 
\hline
B1642$-$03 & & $\surd$ & low & & $\surd$ & \\
B1702$-$19 & & $\surd$ & medium & & $\surd$ & \\
B1702$-$19(interpulse) & $\surd$ & & high & & & \\
B1737+13 & $\surd$ & & medium & & &  \\
B1737+13 & & $\surd$ & low & & $\surd$ &  \\
B1822$-$09(leading) & $\surd$ & & high & & & \\
B1822$-$09(trailing) & $\surd$ & & low & $\surd$ & & $\surd$ \\
B1929+10 & $\surd$ & & high & & & \\
B0144+59 & $\surd$ & & high & & & \\
B0329+54 & $\surd$ & & low & $\surd$ & & $\surd$ \\
B0329+54 & & $\surd$ & low & $\surd$ & & $\surd$ \\
B0525+21 & $\surd$ & & low & $\surd$ & & $\surd$\\
B0540+23 & $\surd$ & & medium & & & \\
B0740$-$28 & $\surd$ & & high & & & \\
B0919+06 & $\surd$ & & high & & & \\
B2319+60 & $\surd$ & & medium/low & $\surd$ & & $\surd$ \\
B2319+60 & & $\surd$ & medium & & & \\
B2351+61 & $\surd$ & & low & $\surd$ & &$\surd$ \\
\end{tabular}
\caption{A summary of the properties seen in the enhanced average profiles.}
\end{table*}

\section{Discussion and conclusions}

The pulsars observed are listed in Table 1, in some instances with
individual components occupying a separate entry. In columns 2 and 3
of the table we show the classification of the pulsar/component as
core or cone.  Column 4 lists the degree of polarization in three
broad categories, as high, medium or low. In column 5 we indicate if
bimodal PA distributions are present. Columns 6 and 7 denote the
relationship between $|V|$ and $V$. If neither column 6 nor 7 apply,
it is implied that $V > |V|$.

Table 1 shows a clear correlation between the presence of bimodal PA
distributions, low overall polarization and $|V| > V$. This is
expected in the superposed model of OPM: the low polarization
indicates the modes are almost equally strong and, if their amplitudes
are not perfectly covariant, a bimodal distribution of PA will ensue.
The integrated $V$ will be low, although individual pulses can show
high values of $V$ of either handedness.  The fact we observe $|V| >
V$ is a strong indicator of wide, bimodal distributions of $V$ in the
single pulses. However, the low mean $V$ compared to the noise makes
the observed distribution appear unimodal (see Figure
\ref{distributions}). 

The degree of polarization in pulsars is generally higher at 1.4 GHz
than at 4.85 GHz, as was identified by e.g. Manchester
(1971)\nocite{man71b}, Xilouris et al. (1996)\nocite{xkj+96}, von
Hoensbroech, Lesch \& Kunzl (1998\nocite{vlk98}). This was explained
by a superposed OPM model, in which the two modes become equally
strong with increasing frequency (Karastergiou et al. 2002), who also
showed that the degree of covariance between the OPMs decreases with
increasing frequency. 
Low covariance between the OPMs means that the difference in the
intensities of the two modes can have a value which fluctuates
significantly in time. It is this difference that determines the
observed polarization (linear, circular and PA), therefore, as a
result, broader distributions in linear and circular polarization and
bimodal PAs are observed.
The pulsars in Table 1 support this model: not only does the total
degree of polarization decrease with frequency, but also the high
$|V|$ profiles (i.e. broad $V$ distributions) indicate a low degree of
correlation between the OPMs. However, low covariance is not enough to
explain the high frequency results. A test of the requirement of the
OPM model, that each OPM is associated to a particular sense of $V$,
fails for the pulsars that show bimodal PA distributions. The
behaviour seen in these pulsars resembles that of PSR B1133+16, which
strongly suggests that the above OPM model needs some adjustment. This
can be achieved by a statistical model where the sign of $V$ is set
randomly as in the model we have presented here, but the underlying
physical reasons remain unclear. A possible explanation could arise by
considering the altitude from the pulsar surface at which the
polarization characteristics are set. At lower heights one may expect
that the regions of positive and negative charge fluctuate on short
time scales, generating greater randomness in the polarization
properties of the higher frequency emission. As the plasma streams
outwards, it becomes more homogeneous, so the lower frequency
polarization properties fluctuate less. It is, however, very important
to stress that the sign-changing process is only present when bimodal
PA distributions are observed, suggesting a common origin of these two
phenomena.

In the pulsars of our sample that show no evidence of bimodal PA
distributions, $V\geq |V|$. The relationship between $V$ and $|V|$
therefore appears to be related to the type of PA distributions,
regardless of the classification of the component. This is highlighted
in PSR B0329+54, where both the core and cone components show bimodal
PA distributions and $|V|>V$. In our sample, bimodal PA distributions
are predominantly observed in cone components. We therefore suggest
that the low degree of average circular polarization in cone emission
is not intrinsic to the emission mechanism, but rather due to broad,
symmetrical distributions of $V$ which are correlated with bimodal PA
distributions. Pulsars with the same characteristics were also
identified in KJMLE. In particular, PSRs B0950+08, B1133+16 and
2020+28 show $|V|>V$ together with bimodal PA distributions. All three
of these pulsars exhibit the aforementioned OPM frequency dependence.

We have identified three pulsars in which $V$ swings from one sign to
the other in the average profile and $|V|=0$ at the same phase bin as
$V=0$. KJMLE also identified 3 such pulsars (PSRs B1508+55, B1933+16
and B2111+46) and argued that these phases correspond to constant, very
low circular polarization. However, such swings in the handedness of
$V$ are also often seen in cone components in the single pulses. It
would therefore be interesting to identify a pulsar that has a cone
component, no OPM jumps and a $V$ profile that shows this
signature. We predict that $|V|$ will behave in the same way as it
does in the core components, i.e. there will be a phase bin where
$V=|V|=0$. This will demonstrate that the $V$ swing is not a
distinguishing feature between cone and core emission and it is for
other reasons, such as jittering in pulse-phase of cone components,
that such swings are usually only seen in the average profiles of core
components.

In conclusion, forming the integrated profile of $|V|$ provides
information not available from the $V$ profile alone. Our main results
are {\bf 1)} that pulse components with bimodal PA distributions also
show wide distributions of $V$, {\bf 2)} that there is little
difference in the circular polarization between core and cone
components except that the swing of $V$ seen in core components is
generally maintained in the single pulses and {\bf 3)} that the degree
of polarization and the covariance between the modes are less in the
pulsars observed at the highest of the two frequencies, as expected by
previous studies, while at the same time the handedness of $V$ is less
correlated with the dominant OPM.

\section*{Acknowledgments}

We thank Dipanjan Mitra for useful ideas and help with the
observations in Effelsberg. SJ is grateful to Richard Wielebinski for
hosting his visit to the MPIfR in Bonn.
 
\label{lastpage}
\bibliographystyle{mn2e}
\bibliography{journals,modrefs,psrrefs}
\end{document}